\documentclass[%
 reprint,%
 amssymb, amsmath,%
 aip,
]{revtex4-1}
\usepackage{graphicx}
\usepackage{dcolumn}
\usepackage[font=scriptsize]{caption}
\usepackage{subcaption}
\usepackage{float}
\usepackage{color}
\usepackage[normalem]{ulem}
\usepackage{booktabs}
\usepackage{url}

\begin{document}

\title{Graphene defect formation by extreme ultraviolet generated photoelectrons} 

\author{A. Gao}
\email{a.gao@utwente.nl.}
 \affiliation{FOM-Dutch Institute for Fundamental Energy Research, Edisonbaan 14,3439 MN Nieuwegein, the Netherlands.}
 \affiliation{XUV Optics Group, MESA+ Institute for Nanotechnology, PO Box 217, University of Twente, 7500 AE, Enschede, the Netherlands}

\author{C.J. Lee}
  \affiliation{FOM-Dutch Institute for Fundamental Energy Research, Edisonbaan 14,3439 MN Nieuwegein, the Netherlands.}
 \affiliation{XUV Optics Group, MESA+ Institute for Nanotechnology, PO Box 217, University of Twente, 7500 AE, Enschede, the Netherlands}
\author{F. Bijkerk}
 \affiliation{FOM-Dutch Institute for Fundamental Energy Research, Edisonbaan 14,3439 MN Nieuwegein, the Netherlands.}
 \affiliation{XUV Optics Group, MESA+ Institute for Nanotechnology, PO Box 217, University of Twente, 7500 AE, Enschede, the Netherlands}
 \date{\today}

\begin{abstract}
We have studied the effect of photoelectrons on defect formation in graphene during extreme ultraviolet~(EUV) irradiation. Assuming the major role of these low energy electrons, we have mimicked the process by using low energy primary electrons. Graphene is irradiated by an electron beam with energy lower than 80~eV. After e-beam irradiation, it is found that the D peak, I(D), appears in the Raman spectrum, indicating defect formation in graphene. The evolution of I(D)/I(G) follows the amorphization trajectory with increasing irradiation dose, indicating that graphene goes through a transformation from microcrystalline to nanocrystalline and then further to amorphous carbon. Further, irradiation of graphene with increased water partial pressure does not significantly change the Raman spectra, which suggests that, in the extremely low energy range, e-beam induced chemical reactions between residual water and graphene is not the dominant mechanism driving defect formation in graphene. Single layer graphene, partially suspended over holes was irradiated with EUV radiation. By comparing with the Raman results from e-beam irradiation, it is concluded that the photoelectrons, especially those from the valence band, contribute to defect formation in graphene during irradiation.  
\end{abstract}


\maketitle 
\thispagestyle{plain}
\pagestyle{plain}

\section{\label{sec:intro}Introduction}

Graphene, a two-dimensional hexagonal packed sheet of carbon atoms, has attracted a lot of attention from different research fields due to its unique physical and chemical properties~\cite{geim2007rise,geim2009graphene,novoselov2005two,zhang2005experimental,han2007energy,bolotin2008ultrahigh,lee2008measurement,bunch2007electromechanical}. However, defects in graphene may substantially influence the performance of graphene-based devices and materials. Irradiation of graphene with energetic particles, such as electrons, ions or photons, is known to generate defects in graphene~\cite{gao2013extreme,zhou2009instability,teweldebrhan2009modification,iqbal2012effect,xu2010monitoring,meyer2012accurate,tao2013modification,lucchese2010quantifying}. In the case of electron irradiation, defect formation in graphene has been extensively studied using transmission electron microscopy~(TEM)~\cite{meyer2012accurate}. In these studies, the same electron beam is used both to irradiate and image graphene, therefore, formation of defects is monitored in situ at atomic resolution. The electron beam energy in TEM is typically higher than the carbon atom displacement threshold in the graphene structure~(80-100~keV)~\cite{banhart2010structural}, leading to vacancy type defects~\cite{meyer2012accurate}. Electron irradiation of graphene with electron energies lower than the displacement threshold has also been reported. Iqbal and Teweldebrhan reported separately that defects appeared in graphene after irradiation with a 20~keV electron beam~\cite{teweldebrhan2009modification,iqbal2012effect}. Furthermore, based on the evolution of D and G peak in Raman spectroscopy, they suggested that graphene went through a transition from crystalline to nanocrystalline and, finally, to amorphous carbon. Irradiation of graphene with energetic photons has also been studied, since graphene-based devices may be used in the presence of ionizing radiation~\cite{gao2013extreme,zhou2009instability}. Zhou reported that soft x-rays can easily break the sp$^2$ bond structure and form defects in graphene that is weakly bound to a substrate~\cite{zhou2009instability}. In their study, exfoliated bi-layer graphene, partially suspended over a trench with a depth of a few micrometers, was also exposed to X-ray radiation. Their analysis showed very similar D peak intensities for the Raman spectra of both the suspended and unsuspended regions. Therefore, it was concluded that defect formation was intrinsic to the graphene and not relevant to the substrate or any gases trapped in the trench. 

The above mentioned studies~\cite{zhou2009instability,teweldebrhan2009modification,iqbal2012effect,xu2010monitoring,meyer2012accurate,tao2013modification,lucchese2010quantifying} on the effects of irradiation on graphene are typically done with graphene on a substrate. The observed defect generation in graphene is usually attributed to the primary irradiation, and the role of photoelectrons or secondary electrons, emitted from the substrate in response to the primary irradiation, has not been discussed in detail. However, in surface photochemistry, secondary electrons are considered to be the dominant factor responsible for surface processes~\cite{madey2006surface,zhou1991photochemistry}. In the study of Zhou and coworkers~\cite{zhou2009instability} it was not possible to discuss the effect of the secondary electrons~(photoelectrons) on defect generation in graphene in a quantitative way. This is because the secondary electron~(photoelectron) yield of the SiO$_2$ substrate is unknown.

In this letter, we study graphene defect generation due to direct exposure to electrons with energies that are typical for photoelectrons.  Furthermore, by increasing the partial pressure of water in the chamber, we show that defects do not arise from electron-induced surface chemistry. By comparing the rate at which defects are generated by direct, low energy electrons and EUV generated electrons, we show that the EUV-induced photoelectrons, especially those from the valence band, contribute to defect formation in graphene during irradiation.

\begin{figure*}[htb]
\centering
                \includegraphics[width=\textwidth]{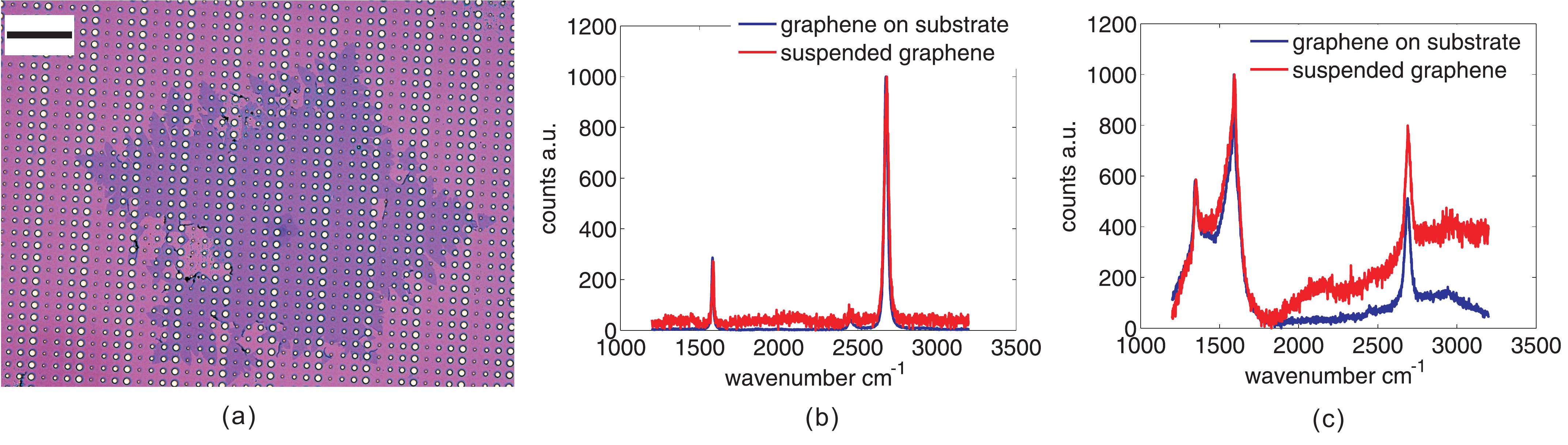}
\caption{\label{fig:spectra} (color online) (a) optical image of graphene partially suspended on the holes on SiO$_2$/Si substrate. The darker purple area indicates where the graphene is. The scale bar in the image is 50~$\mu$m. (b) The Raman spectra for the graphene suspended and supported regions before EUV irradiation. (c) Raman spectra for the graphene suspended on a 4~$\mu$m hole and supported regions after irradiation. All the Raman spectra have been normalized.}
\end{figure*}

\section{\label{sec:exp}Experiments}
Single layer graphene samples were obtained from Graphene Master and Graphene Supermarket. In both cases, the graphene was grown by chemical vapor deposition on copper and transferred to a SiO$_2$/Si substrate with a 285~nm thick layer of SiO$_2$. The samples from Graphene Master were placed on a 5~mm square substrate that had a two dimensional array of holes etched into it. The diameter of the holes varied from 2~$\mu$m to 5~$\mu$m and had a depth of 300~nm, so that the transferred graphene was partially suspended. The samples from Graphene Supermarket were transferred to a 10~mm square, unstructured substrate for low energy electron beam studies.  

EUV exposures were performed using radiation from a Xe plasma discharge source~(Philips EUV Alpha Source 2) with a repetition rate of 1000~Hz. \textcolor{red}{After passing through a Si/Mo/Zr thin membrane filter, the spectrum of the EUV radiation has three emission lines at 11~nm, 13.5~nm, and 15~nm, with bandwidth of about 1~nm for each line. The out-of-band deep UV radiation is less than 3\% of the transmitted power~\cite{klosner1998intense,Sjmaenok2012,Banine2008}.} The EUV beam profile has a Gaussian distribution with full width half maximum of 3~mm. The peak EUV intensity at the sample surface was estimated to be 5~W/cm$^2$ with a dose of 5~mJ/cm$^2$ per pulse. The base pressure of the EUV exposure chamber was 1x10$^{-8}$~mbar, which increases to 5x10$^{-7}$~mbar during irradiation due to a small amount of Xe/Ar gas mixture from the source chamber leaking into the exposure chamber. 

E-beam irradiation was performed with an ELG-2/EGPS-1022 electron gun~(Kimball physics). The electron energy was varied from 3.7~eV to 80~eV, while the electron dose was controlled by varying the irradiation time and emission current. The distance between the electron gun and the grounded sample was approximately 25~mm. E-beam exposures were performed at a chamber base pressure of 5x10$^{-9}$~mbar, which increased only when additional background gases were deliberately added. 
Raman spectra were collected with a commercial Raman microspectrometer system~(Renishaw) with an excitation wavelength of 514~nm, a spot size of 1~$\mu$m and an excitation power of 2.5~mW. A home-built Raman spectrometer, based on a 532~nm excitation wavelength, an illumination intensity of 200~W/cm$^2$, and a spectrometer~(Solar Laser system M266) with a resolution of 1~cm$^{-1}$, was also used to collect wide-area Raman maps. The collection optics and pixel size of the detector result in a spatial resolution of 100x100~$\mu$m$^2$ and a field of view of 100~$\mu$m by 1000~$\mu$m. 
\begin{figure*}
\centering
                \includegraphics[width=0.8\textwidth]{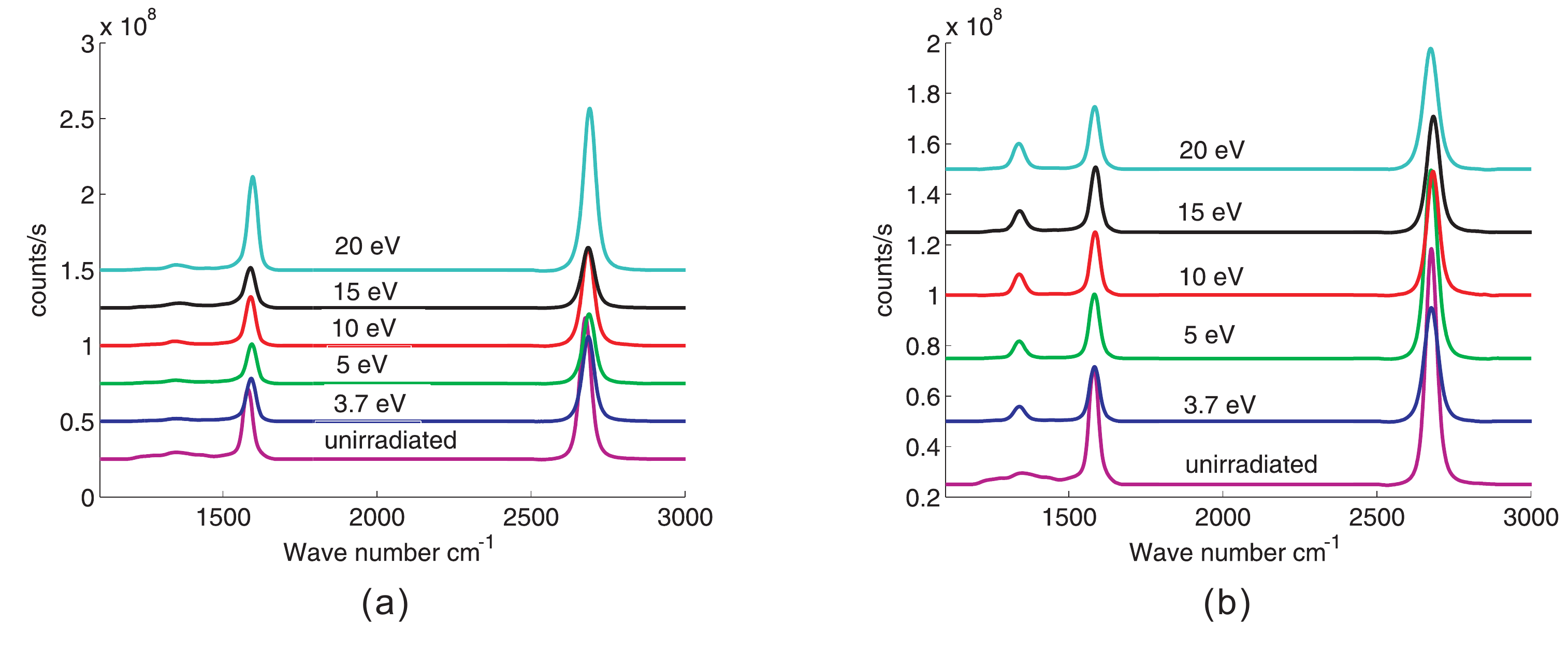}        
        \caption{(color online) Raman spectra of graphene on SiO$_2$/Si~(no holes, all graphene supported on SiO$_2$) irradiated with different e-beam energies. The electron dose (a) 1x10$^{17}$~cm$^{-2}$, and (b) 1x10$^{18}$~cm$^{-2}$~(except for 3.7~eV and 5~eV, which is 1x10$^{19}$~cm$^{-2}$).}\label{fig:ramanelectron}
\end{figure*}

\section{\label{sec:results}Results and Discussion}
Fig.~\ref{fig:spectra}a shows the optical image of single layer graphene partially suspended on the SiO$_2$/Si substrate. The Raman spectra for the graphene suspended and supported regions before EUV irradiation are shown in Fig.~\ref{fig:spectra}b. The I(2D)/I(G) is about 4, and the full width of the 2D peak is about 30~cm$^{-1}$, confirming that the graphene is single layer. After EUV irradiation, the Raman spectra for the graphene suspended and supported regions are plotted in Fig.~\ref{fig:spectra}c. It is clearly shown that in both regions, a D peak, and a fluorescence background appear. The latter is due to EUV induced carbon contamination~\cite{windt1998imd}. For the graphene on a substrate, assuming that the atmosphere was not too clean, then there is fluorescence from carbon on top of the graphene as well as on the bottom of the graphene. But, for the suspended sample, there is an additional signal from carbon at the bottom surface of the hole as well as signal from around the edges of the diffraction-limited spot~(a ring that appears from the point of view of the microscope image plane) to have originated from the diffraction-limited spot at the graphene surface. A simple calculation reveals that this can lead to an enhancement of contributing area of approximately 2, while the fluorescence background is about 2.4 times greater. The paper has been changed to indicate this.The two spectra have approximately the same I(D) but differ in the fluorescence background from 1800~cm$^{-1}$ and higher wavenumbers. \textcolor{red}{For the graphene on a substrate, there is fluorescence from hydrocarbon adsorbed on both sides of graphene. However, for the suspended sample, there is an additional signal from hydrocarbon at the bottom surface of the hole. In addition, the geometry allows for a contribution from a ring on the Si surface that, geometrically, will appear to have originated from the diffraction-limited spot on the graphene surface. These additional contributions can lead to an enhancement of the fluorescence background of about 2.4 times.} The same I(D) indicates that the defect density is the same in both suspended and supported regions. 

The experimental results here give rise to an interesting conclusion: either the photoelectrons emitted from the substrate do not generate any defects in graphene, or the photoelectrons emitted from both regions~(graphene on SiO$_2$ graphene suspended over Si) result in the same defect density in graphene, despite having vastly different photoelectron yields.  It is likely that the photoelectrons, which typically have an energy spectrum with a maximum near the work function of the material~(\textless10~eV) from which they are emitted, do not have sufficient energy to generate defects in graphene.

\textcolor{red}{The photoelectron energy spectrum from Si with native oxide starts at around 2~eV and is sharply peaked at around 2.5~eV, with a full width half maximum of 0.86~eV. At higher energies, the photoelectron yield decays exponentially. Electrons with energies above 20~eV are rarely emitted with an exception at 80-85~eV, corresponding to emission directly from the valence band. The flux of electrons within the energy range of 80-85~eV is approximately 3\% of the total dose.} To test if photoelectrons can damage graphene, graphene samples were irradiated using the low energy electron gun. Fig.~\ref{fig:ramanelectron} shows the Raman spectra of graphene on an unstructured SiO$_2$/Si after irradiation of electrons with different energies . In Fig.~\ref{fig:ramanelectron}a, where the electron dose is about 1x10$^{17}$~cm$^{-2}$, the Raman spectra of the irradiated graphene samples are almost identical to the unirradiated samples, with no clear D peak. This indicates that no detectable defects were generated in graphene during e-beam irradiation. However, as the electron dose increases to 1x10$^{18}$~cm$^{-2}$, and beyond, as shown in Fig.~\ref{fig:ramanelectron}b, all the irradiated samples show a relatively small but clear D peak in their Raman spectra, confirming defect formation in graphene during irradiation.

\begin{figure*}
\centering
                \includegraphics[width=\textwidth]{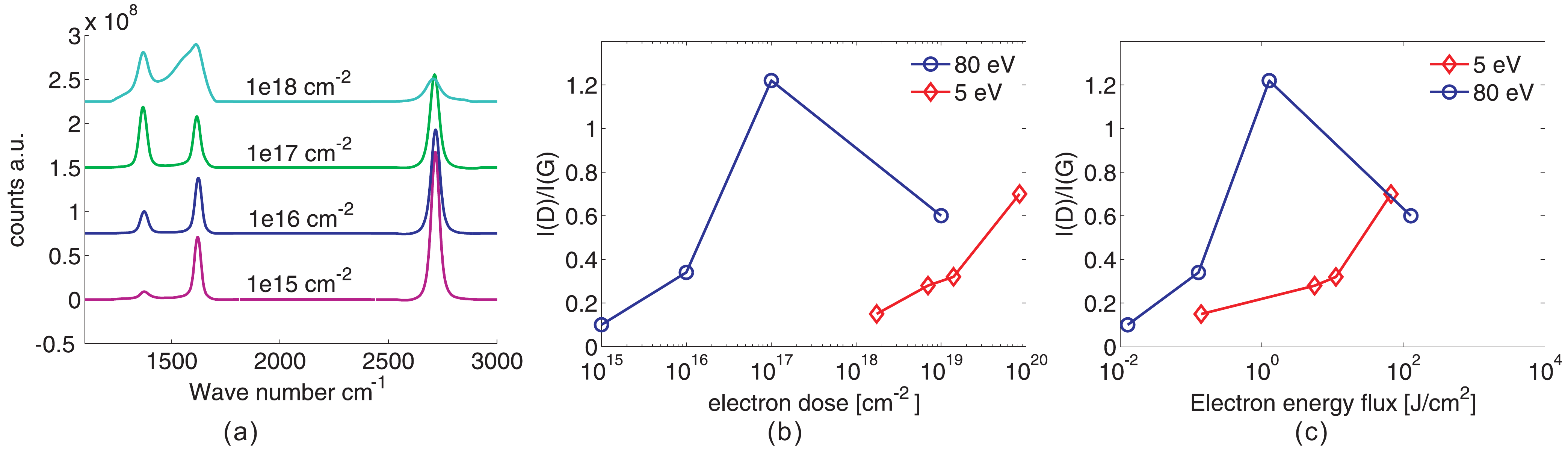}
                \caption{(color online) (a) Raman spectra for graphene samples on SiO$_2$/Si~(no holes, all graphene supported on SiO$_2$) irradiated under different dosages of electrons with energy of 80~eV. (b) I(D)/I(G) ratio versus the electrons dose. (c) I(D)/I(G) ratio versus the electron energy times electron dose. }\label{fig:doseraman}
\end{figure*}

Graphene samples irradiated by electrons with an energy of 80~eV were also examined. The photoelectron energy spectrum of Si has a small peak at 80~eV, due to emission from the valence band under EUV~(92~eV) irradiation. As a result, the photoelectron flux at 80~eV is much greater than the flux at energies between 80 and 20~eV and should be investigated. The Raman spectra of the irradiated samples are shown in Fig.~\ref{fig:doseraman}a. From the Raman spectra, it can be seen that a D peak appears, even at very low dose, indicating 80~eV electrons generate defects in graphene more efficiently. The I(D)/I(G) ratio as a function of the electron dose is plotted in Fig.~\ref{fig:doseraman}b. The I(D)/I(G) ratio first increases to a maximum and then falls with increasing electron dose. This behavior follows the amorphization trajectory in irradiated carbon material proposed by Ferrari~\cite{ferrari2000interpretation}. Electrons first cause local defects in graphene, reducing the long-range order. Thus (micro)~crystalline graphene transforms to nanocrystalline graphene. As the defects accumulate, the nanocrystalline graphene becomes more disordered, until it must be considered to be amorphous sp$^2$ carbon. Note that the I(D)/I(G) ratio as a function of dose for 5~eV electrons is also plotted in Fig.~\ref{fig:doseraman}b and appears to be following the same trajectory, though requiring a larger dose. \textcolor{red}{It should also be noted, however, that the damage is not simply a function of the energy deposited in the sample, as can be seen in Fig.~\ref{fig:doseraman}c. This is because different defect types require different activation energies. Furthermore, the cross section for each defect formation process is likely to be a function of the electron energy.}

\begin{figure}
                \centering
                \includegraphics[width=0.4\textwidth]{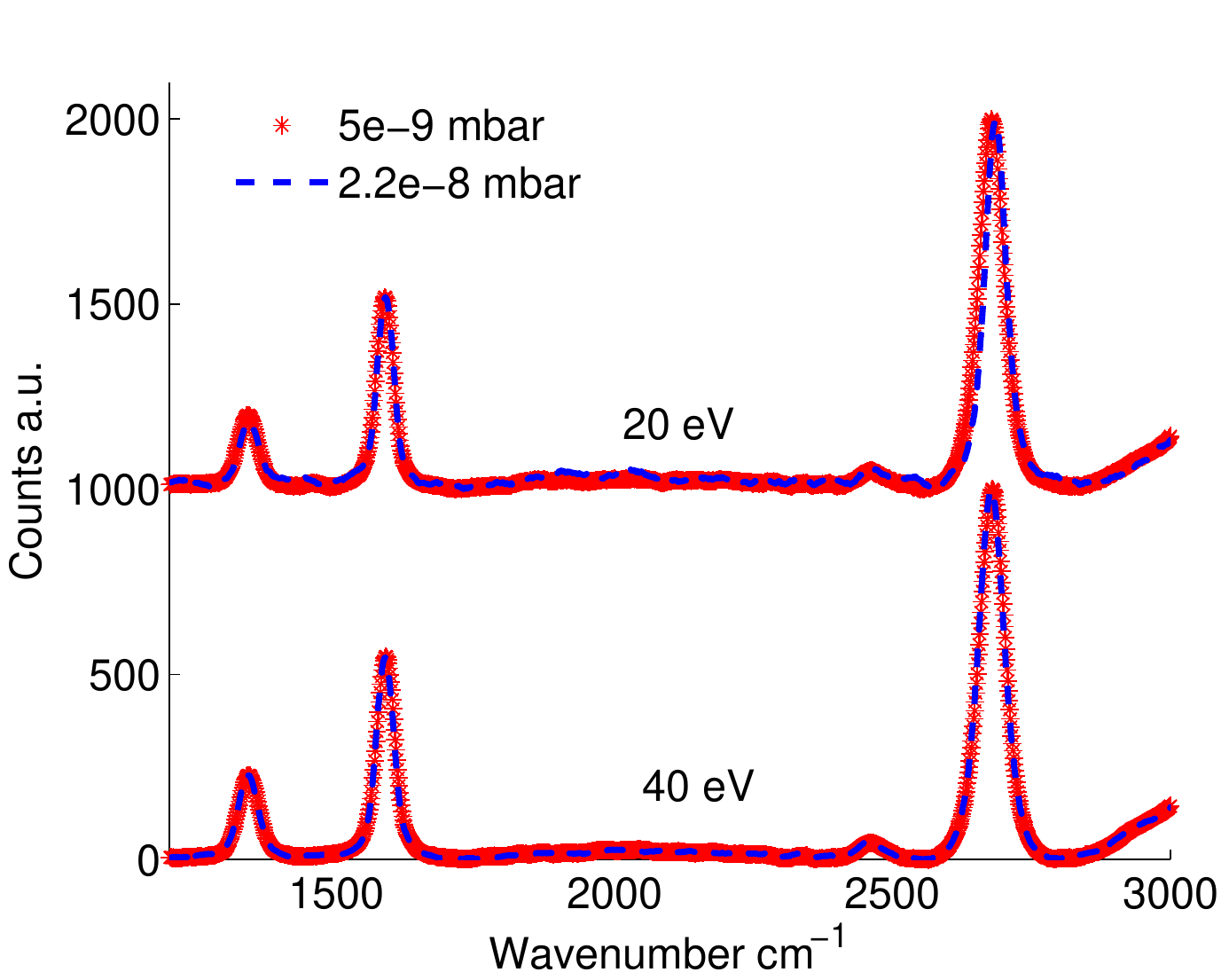}
        \caption{(color online) Raman spectra for graphene on SiO$_2$/Si irradiated by 20 and 40~eV in different vacuum conditions.}\label{fig:water}
\end{figure}

The presence of residual water vapor in the vacuum chamber is known to result in graphene oxidation when exposed to 100~keV electron irradiation~\cite{meyer2012accurate,yuzvinsky2005precision}. It is, therefore, possible, that the observed increase in defects is due to electron-induced chemistry. To test this, graphene was irradiated with 20~eV and 40~eV electrons at two different background water partial pressures. Under normal operating conditions, the main residual gas in the chamber is water, at a maximum pressure of 5x10$^{-9}$~mbar~(in reality it is less, since this is the total chamber pressure). The background water pressure was increased by leaking water into the chamber until the pressure was 2.2x10$^{-8}$~mbar. Note that higher pressures cannot be used because the electron gun only works at pressures below 1x10$^{-7}$~mbar. Fig.~\ref{fig:water} shows the Raman spectra of the graphene samples irradiated by 20~eV and 40~eV electrons at two different chamber pressures. The spectra are almost identical, meaning that, in the extremely low energy range of electron irradiation, the electron flux does not initiate chemical reactions between residual water and graphene at a measurable rate. Therefore, oxidation is not the dominant mechanism for defect formation in graphene. \textcolor{red}{Yuzvinsky \textit{et al} also reported that electron beam induced damage to carbon nanotube was closely related with the water partial pressure~\cite{yuzvinsky2005precision}. In their experiments, no damage was observed for experiments with water partial pressure below 2 x $10^{-6}$~Torr with electron energy of 1~keV.} Furthermore, in Fig.~\ref{fig:ramanelectron}b it is shown that irradiation with 3.7~eV electrons is sufficient to initiate defects in graphene, which is lower than the bond energy of O-H bonds in water (about 4.8~eV) and the ionization energy of water~(about 12.6~eV)~\cite{page1988high}.

As mentioned in the introduction, vacancy type defects in graphene, require an electron energy of 80-100~keV. It is also reported that the Stone-Wales type of defect requires an electron energy of approximately 25~keV~\cite{banhart2010structural}. Since the energy of electrons in this study is far below these values, neither vacancy nor Stone-Wales type of defects are expected here. Krauss et al reported disassembly of a graphene single crystal into a nanocrystalline network induced by 488~nm~(2.54~eV) laser irradiation~\cite{krauss2009laser}. They concluded that the disassembly process is due to two-photon induced breaking of sp$^2$ carbon-carbon bonds. The bond enthalpy for carbon-carbon single bond and double bonds is 3.6~eV and 6.14~eV separately~\cite{atkins}. The carbon single bond energy is about the same as the lowest electron beam energy in our experiments. We conclude, therefore, that defects are due to breaking sp$^2$ and, thus, forming sp$^3$ bonds. As a result, smaller sub-crystal structures form~(nanocrystalline graphene).

Now it is possible to discuss the results in Fig.~\ref{fig:spectra}. The photoelectron yield under EUV~(92~eV) irradiation from a Si surface with native oxide is about 0.017 electrons/photon~\cite{personal}. \textcolor{red}{The natural oxide layer in the holes is not thick enough to prevent photoelectron emission from the underlying Si.} Although there is no published data on the photoelectron yield from SiO$_2$, it was estimated to be 0.001~\cite{unpub}.  For 30~min exposure with an EUV intensity of 5~W/cm$^2$, the total dose of photoelectrons ejected from the silicon surface is about 1x10$^{19}$~cm$^{-2}$, and from SiO$_2$ surface is 6x10$^{17}$~cm$^{-2}$~(assuming the yield is 0.001). According to the data in Fig.~\ref{fig:ramanelectron} and Fig.~\ref{fig:doseraman}, the graphene sitting directly on the SiO$_2$ substrate is exposed to an electron dose which is unlikely to lead to a detectable D peak with an exception of the valence band electrons with energy at around 80~eV, which will contribute an I(D)/I(G) of 0.15. On the other hand, the high photoelectron yield of the Si surface should result in an increase in defects, corresponding to an increase of I(D)/I(G)~= 1.2 relative to the unsuspended graphene. \textcolor{red}{The discrepancy between the damage prediction and experimental observation can be explained that the flux of photoelectrons to the graphene is reduced by the experimental conditions.} It has been shown that graphene, suspended over trenches and holes, is able to trap gas at atmospheric pressure~\cite{kitt2013graphene}. The graphene membrane was transferred onto the SiO$_2$ substrate under atmospheric conditions, therefore, it is possible that the pressure in the hole is approximately 1 bar. Under these conditions, the photoelectrons are likely to scatter given the fact that its mean free path is comparable with the height of the hole. From the Raman spectra in Fig.~\ref{fig:spectra}, we also observed that the fluorescence background, due to hydrocarbon deposition~(on graphene and/or Si), was substantially stronger in the suspended regions. This indicates that an amorphous carbon layer may be shielding the graphene from the photoelectron flux.

It is also interesting to compare our observations to those from electron and EUV irradiation of other surfaces, such as ruthenium~\cite{madey2006surface}. In the case of metals, the dominant form of degradation is due to oxidation~(provided residual hydrocarbons are under control). Published data show that the low energy secondary electrons are primarily responsible for the dissociation of water, leading to the surface and subsurface oxidizing~\cite{madey2006surface}. This is in stark contrast to our results, which indicate that the photoelectrons do not promote oxidation, and that the graphene damage is limited to direct processes, such as sp$^2$ bond breaking.

\section{\label{sec:sum} Conclusion}
We have studied the effect of photoelectrons from a substrate on defect formation in graphene during EUV irradiation. Experiments show that extremely low (less than 80 eV) energy electrons will lead to defect formation in graphene if it is irradiated with sufficient dose. The electrons excited directly from the valence band are more efficient in defect formation than the photoelectrons with lower energies~(less than 20~eV). The process of the damage to graphene follows the amorphization trajectory with increasing irradiation dose, indicating that graphene goes through a transformation to nanocrystalline and then further to amorphous carbon. Furthermore, irradiation of graphene with different water partial pressures show similar Raman spectra, which suggests that, in the extremely low energy range, e-beam induced chemical reactions between residual water and graphene is not the dominant effect in defect formation in graphene. These results indicate a different degradation process compared to the EUV induced oxidation of metallic surfaces, namely photo-induced electrons break sp$^2$ bonds and, thus, lead to graphene degradation during EUV radiation. These findings are of relevance for protective top layers on EUV reflecting mirrors in applications, such as EUV lithography.

\begin{acknowledgments}
The authors would like to thank Goran Milinkovic, Luc Stevens, and John de Kuster for the help with sample preparation and experimental measurements, Ren¨¦ Vervuurt and Jan-Willem Weber for the micro-Raman measurements. This work is part of the research programme Controlling photon and plasma induced processes at EUV optical surfaces (CP3E) of the Stichting voor Fundamenteel Onderzoek der Materie (FOM) with financial support from the Nederlandse Organisatie voor Wetenschappelijk Onderzoek (NWO).  The CP3E programme is co-financed by Carl Zeiss SMT and ASML, and the AgentschapNL through the EXEPT programme.
\end{acknowledgments}


%

\end{document}